\documentclass[preprint,12pt,number,1p]{elsarticle}
\usepackage{graphicx,color}
\usepackage{amsmath}

  \newcommand {\nc} {\newcommand}
  \nc {\beq} {\begin{eqnarray}}
  \nc {\eeq} {\nonumber \end{eqnarray}}
  \nc {\eeqn}[1] {\label {#1} \end{eqnarray}}
  \nc {\mrm} [1] {\mathrm{#1}}
  \nc {\bce}{\begin{center}}
  \nc {\ece} {\end{center}}
  \nc {\ex} [1] {\ensuremath{^{#1}}}
  \nc {\btrs} {\begin{tabular*}}
  \nc {\etrs} {\end{tabular*}}
  \nc {\btr} {\begin{tabular}}
  \nc {\etr} {\end{tabular}}
  \nc {\half} {\mbox{$\frac{1}{2}$}}
  \nc {\thal} {\mbox{$\frac{3}{2}$}}
  \nc {\fial} {\mbox{$\frac{5}{2}$}}
  \nc {\etal} {\emph{et al.\ }}
  \nc {\ve} [1] {\mbox{\boldmath $#1$}}
  \nc {\ves} [1] {\mbox{\boldmath ${\scriptstyle #1}$}}
  \nc {\eq} [1] {(\ref{#1})}
  \nc {\Eq} [1] {Eq.~(\ref{#1})}
  \nc {\Ref} [1] {Ref.~\cite{#1}}
  \nc {\Refc} [2] {Refs.~\cite[#1]{#2}}
  \nc {\Sec} [1] {Sec.~\ref{#1}}
  \nc {\chap} [1] {Chapter~\ref{#1}}
  \nc {\anx} [1] {Appendix~\ref{#1}}
  \nc {\tbl} [1] {Table~\ref{#1}}
  \nc {\fig} [1] {Fig.~\ref{#1}}
  \nc {\Sch} {Schr\"odinger }
  \nc {\flim} [2] {\mathop{\longrightarrow}\limits_{{#1}\rightarrow{#2}}}

\begin{document}

\author[msu,ulb]{P. Capel\fnref{fn1}\corref{cor}}
\ead{capel@nscl.msu.edu}
\author[usp]{M. S. Hussein}
\ead{hussein@if.usp.br}
\author[ulb]{D. Baye}
\ead{dbaye@ulb.ac.be}
\address[msu]{National Superconducting Cyclotron Laboratory, Michigan State University, East Lansing MI 48824, USA}
\address[ulb]{Physique Quantique, CP 165/82, and Physique Nucl\'eaire Th\'eorique et Physique Math\'ematique, CP229,
Universit\'e Libre de Bruxelles (ULB), B-1050 Brussels, Belgium}
\address[usp]{Instituto de F\'{i}sica, Universidade de S\~{a}o Paulo
C.P. 66318, 05315-970 S\~{a}o Paulo, S.P., Brazil}

\fntext[fn1]{Postdoctoral researcher of the F.R.S.-FNRS}
\cortext[cor]{Corresponding author}
\title{Influence of the halo upon angular distributions for elastic scattering and breakup}

\begin{abstract}
The angular distributions for elastic scattering and breakup of halo
nuclei are analysed using a near-side/far-side decomposition
within the framework of the dynamical eikonal approximation.
This analysis is performed for \ex{11}Be impinging on Pb at 69~MeV/nucleon.
These distributions exhibit very similar features.
In particular they are both
near-side dominated, as expected from Coulomb-dominated reactions.
The general shape of these distributions is sensitive mostly to
the projectile-target interactions,
but is also affected by the extension of the halo.
This suggests the elastic scattering not to be affected by a loss of
flux towards the breakup channel.
\end{abstract}

\begin{keyword}
Halo nuclei, elastic scattering, breakup,
angular distribution, near/far decomposition
\end{keyword}

\maketitle

\section{Introduction}
Since their discovery, halo nuclei have been the subject of many
theoretical and experimental investigations \cite{Tan96}.
These light exotic nuclei exhibit a peculiar quantal structure
due to their low separation energy of one or two nucleons.
Thanks to this weak binding, these valence nucleons have a significant
probability of presence at large distances
and form a sort of halo around the core of the nucleus \cite{HJ87}.
Since they appear at the edge of the valley of stability,
halo nuclei are short-lived and cannot be studied
through usual spectroscopic techniques.
One usually resorts to reactions
to infer information about their structure.
Elastic scattering, being a peripheral process,
is sensitive mostly to the tail of the wave functions of the colliding nuclei.
It may thus be an interesting probe of 
these exotic nuclei.
Moreover, elastic scattering is likely to be affected
by reactions, which, like breakup, are significantly enhanced by
the weak binding of these nuclei.
For that reason, efforts have been made to investigate the effects
of the halo upon elastic scattering around the Coulomb
barrier \cite{Dip04,Dip10}. These works have shown a significant
reduction of the elastic-scattering cross section
at large angles compared to stable nuclei.
This reduction is interpreted as
a loss of flux from the elastic channel towards reaction channels, like
breakup or transfer.

Angular distributions for both elastic scattering and breakup
are also studied at intermediate energies \cite{Kik97,Fuk04}.
In \Ref{Kik97}, such observables have been measured
to evaluate the E2 contribution to the Coulomb breakup
of \ex{8}B, which is of astrophysical interest.
In \Ref{Fuk04}, angular distributions for the
elastic scattering and breakup of \ex{11}Be have been measured
to investigate the structure of this one-neutron halo nucleus.
Both experiments have been successfully reproduced within the
dynamical eikonal approximation (DEA) \cite{BCG05,GBC06,GCB07},
which coherently describes both elastic scattering and breakup.
Interestingly, the theoretical analysis of the
\ex{8}B Coulomb breakup has shown significant nuclear-Coulomb
interferences
that may convey information about the structure of the projectile and/or
its interaction with the target \cite{GCB07}.
To better understand these features and
the interplay between the halo structure
and the reaction mechanism, we follow the idea of \Ref{CIH85} and
perform a near-side/far-side (N/F) analysis of the DEA angular distributions 
(see also \Ref{HM84} for a review).
To get a better insight into the influence of breakup
onto elastic scattering, we analyse the angular distributions
for both processes simultaneously.

We first give a brief account of the DEA
and summarise how angular distributions are calculated within this model.
We then derive the N/F decomposition of the elastic-scattering
cross section 
and extend that decomposition to angular distributions for elastic breakup.
Based on these theoretical developments, we analyse the elastic scattering
and breakup of \ex{11}Be, the archetypal one-neutron
halo nucleus, on lead at 69~MeV/nucleon,
which correspond to the conditions of the RIKEN experiment of \Ref{Fuk04}.

\section{Theoretical considerations}

In a nutshell, the DEA \cite{BCG05,GBC06}
hinges on using a three-body description
of the projectile-target system:
The projectile $P$ is seen as a fragment $f$ loosely-bound to a core $c$.
This two-cluster structure is described by the internal Hamiltonian 
$H_0$, which depends on the $c$-$f$ relative coordinate $\ve{r}$
and in which the $c$-$f$ interaction is simulated by a real potential.
The target $T$ is seen as a structureless particle, and its interactions
with the core and the fragment are simulated by the optical potentials
$V_{cT}$ and $V_{fT}$.
In the DEA, the resulting three-body \Sch equation
is solved using the eikonal ansatz for the wave function:
$\Psi(\ve{r},\ve{R})= e^{iKZ} \widehat{\Psi}(\ve{r},\ve{R})$ \cite{Glauber},
$Z$ being the component along the beam axis of
$\ve{R}$, the $P$-$T$ relative coordinate,
and $\hbar K$ the initial momentum of the $P$-$T$ relative motion.
At sufficiently high energy, $\widehat{\Psi}$ varies smoothly with $\ve{R}$,
and its second-order derivative in $\ve{R}$ can be neglected in front of
its first-order derivative \cite{Glauber}.
This leads to the DEA equation \cite{BCG05,GBC06}
\beq
i\hbar v \frac{\partial}{\partial Z}\widehat\Psi(\ve{r},\ve{R}) = (H_0 + V_{cT} + V_{fT} - E_0) \widehat\Psi(\ve{r},\ve{R}),
\eeqn{e3}
where $v$ is the initial $P$ velocity,
and $E_0$ the (negative) energy of the projectile
ground-state $\phi_{l_0j_0m_0}$.
The quantum numbers $l_0$, $j_0$ and $m_0$ correspond to
the $c$-$f$ orbital momentum,
the projectile total angular momentum and its projection, respectively.
The spin of the core being neglected, $j_0$ is obtained from
the coupling of the fragment spin to the orbital momentum.
For each value of $\ve{b}$, the transverse component of $\ve{R}$,
\Eq{e3} is solved numerically with the initial condition
that the projectile is in its ground state:
$\widehat\Psi^{(m_0)}(\ve{r},\ve{b},Z) \flim{Z}{-\infty} \phi_{l_0j_0m_0}(\ve{r}).$

The transition matrix element for elastic scattering 
from projection $m_0$ to $m'_0$ in the direction
$\ve{K}=(K,\theta,\varphi)$ can be obtained
from the solutions of \eq{e3} \cite{GBC06}
\beq
T_{\rm el}=2\pi\hbar v i^{1-|m_0-m'_0|}e^{i(m_0-m'_0)\varphi}
\int_0^\infty b db J_{|m_0-m'_0|}(qb)S_{{\rm el}, m'_0}^{(m_0)}(b),
\eeqn{e5}
where $J_\mu$ is a Bessel function \cite{AS70},
$q=2K\sin\theta/2$ is the transferred momentum, and
the elastic-scattering amplitudes read
\beq
S_{{\rm el}, m'_0}^{(m_0)}(b)=\langle \phi_{l_0j_0m'_0}(\ve{r})|
\widehat\Psi^{(m_0)}(\ve{r},b\ve{\hat X},Z)\rangle_{Z\rightarrow +\infty}
-\delta_{m'_0m_0}.
\eeqn{e6}

Following \Ref{CIH85}, we divide the $T_{\rm el}$
matrix element \eq{e5} into its near (N) and far (F) sides
by decomposing the Bessel function into Hankel functions
\cite {AS70}:
$J_\mu(z)=\frac{1}{2}\left[H_\mu^{(2)}(z)+H_\mu^{(1)}(z)\right]$.
The N side corresponds to the expression \eq{e5} in which
$J_{|m_0-m'_0|}(qb)$ is replaced by $H_{|m_0-m'_0|}^{(2)}(qb)/2$,
and the F side to the same expression with $H_{|m_0-m'_0|}^{(1)}(qb)/2$
instead.
The N and F sides are related to positive and negative scattering angles,
as can be understood from the asymptotic behaviour
of the Hankel functions
\beq
H_\mu^{\tiny \left(\!\!\!\!\begin{tabular}{c}
{1}\\
{2}
\end{tabular}
\!\!\!\!\right)}(z)\flim{z}{\infty}\sqrt{\frac{2}{\pi z}}
e^{\pm i(z-\mu\pi/2-\pi/4)}.
\eeqn{e7b}
Remembering that $q=2K\sin\theta/2\approx K\theta$, one sees that
the N side arises from positive deflection, while
the F side corresponds to negative deflection \cite{CIH85}.
The former can thus be interpreted as the contribution of repulsive
forces in the $P$-$T$ scattering, and the latter as resulting from
the attractive parts of the $P$-$T$ interactions.
As shown in Refs.~\cite{CIH85,HM84}, this decomposition is a powerful tool
to analyse reaction mechanisms.

Similarly, a $T$ matrix can be defined for the dissociation of the
projectile \cite{GBC06}. For a final $P$-$T$ wave vector
$\ve{K'}=(K',\theta,\varphi)$, it reads
\beq
T_{\rm bu}=2\pi\hbar v\sum_{ljm}
i^{1-|m_0-m|}e^{i(m_0-m)\varphi}
\int_0^\infty b db J_{|m_0-m|}(qb)S_{\ves{k}\nu,ljm}^{(m_0)}(b),
\eeqn{e8}
where the transferred momentum $q\approx 2K\sin\theta/2$
and the breakup amplitude
\beq
S_{\ves{k}\nu,ljm}^{(m_0)}(b)=\langle \chi_{\ves{k}\nu,ljm}^{(-)}(\ve{r})|
\widehat\Psi^{(m_0)}(\ve{r},b\ve{\hat X},Z)\rangle_{Z\rightarrow +\infty}.
\eeqn{e9}
In \Eq{e9}, $\chi_{\ves{k}\nu,ljm}^{(-)}$ is the component in the partial wave
$ljm$ of the incoming distorted scattering wave $\chi_{\ves{k}\nu}^{(-)}$
describing the broken-up projectile
with a $c$-$f$ wave vector $\ve{k}$ and a spin projection $\nu$.
These expressions obtained within the DEA enable us to
generalise the N/F decomposition to angular distributions
for breakup.
Using the same decomposition of the Bessel function in \Eq{e8},
we obtain N and F sides of $T_{\rm bu}$ with the same interpretation.

\section{Elastic scattering}
We first exploit these theoretical developments by
analysing the angular distributions obtained within the DEA for
the elastic scattering of \ex{11}Be on Pb at 69~MeV/nucleon.
This one-neutron halo nucleus is described as a \ex{10}Be core
in its $0^+$ ground state to which a neutron is by 0.5~MeV,
as explained in \Ref{GBC06}.
In that model, the $1/2^+$ ground state
is reproduced considering the halo neutron in the $1s_{1/2}$ orbital.
We use the optical potentials listed in \Ref{GBC06}
to simulate the $P$-$T$ interactions.

\begin{figure}
\center
\includegraphics[width=10cm]{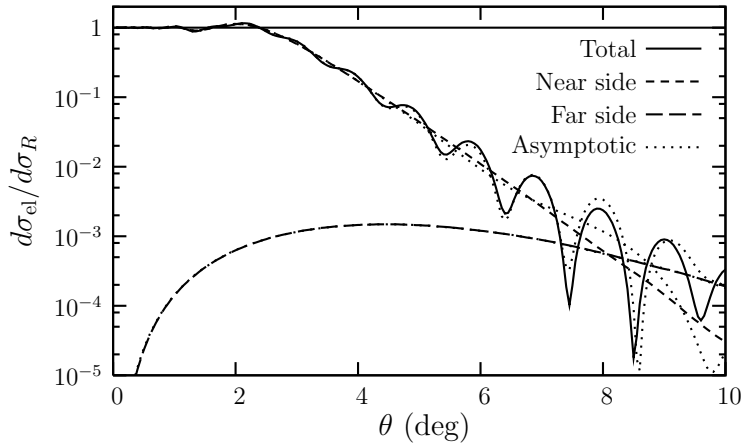}
\caption{Elastic-scattering cross section and its N/F decomposition
for \ex{11}Be impinging on Pb at 69~MeV/nucleon.
The results obtained with the asymptotic expression of 
the Bessel and Hankel functions are shown as dotted lines.
}\label{f1}
\end{figure}

The elastic-scattering cross section,
normalised to the Rutherford cross section,
is plotted in \fig{f1} as a function of the scattering angle
$\theta$ (full line).
At forward angles, it presents a usual rainbow pattern
around the Rutherford cross section.
During its drop beyond $2^{\circ}$, the cross section
exhibits a second oscillatory pattern.
To understand these features, we plot the N and
F sides of that cross section.
As expected for a reaction dominated by the (repulsive) Coulomb interaction,
the elastic-scattering cross section follows its N side at
forward angles. It is only at larger angles, after the N side
has started to drop, that the F side becomes significant
and crosses the N side at about $8^{\circ}$.
This crossover explains the second oscillatory pattern as a N/F interference.

The interpretation of the N/F decomposition as positive/negative
deflection is based on the asymptotic behaviour of the Hankel
functions \eq{e7b}. To assess its validity, we repeat the calculations
using the asymptotic expression of the Bessel and Hankel functions.
These results, displayed as dotted lines in \fig{f1},
are nearly superimposed on the exact expressions,
confirming the N/F interpretation.

\begin{figure}
\center
\includegraphics[width=10cm]{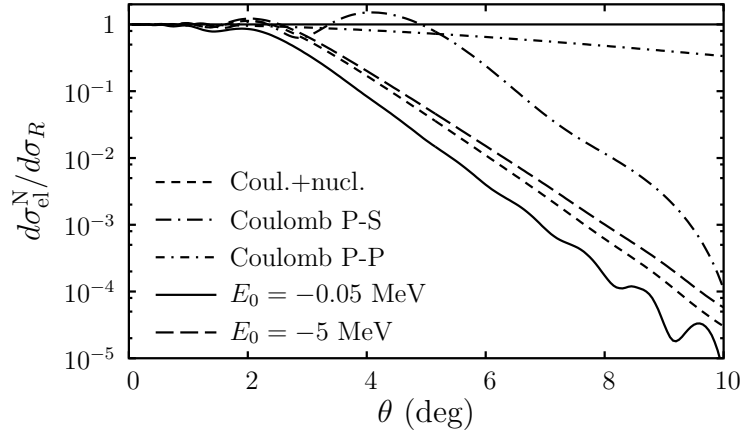}
\caption{Influence of the nature of the $P$-$T$ interactions 
and of the extension of the halo on the N side of
the elastic-scattering cross section
of \ex{11}Be on Pb at 69~MeV/nucleon.
}\label{f2}
\end{figure}

To evaluate the sensitivity of
the pattern of the elastic-scattering cross section
to the $P$-$T$ interactions,
the calculation is repeated considering only 
the Coulomb interaction simulated either by the
point-sphere term of the $c$-$T$ optical potential (P-S)
or a mere point-point potential between
the core and the target (P-P).
Their dominant N sides are displayed in \fig{f2}.
We observe two very different results.
The P-S Coulomb interaction leads to qualitatively
similar features as the full optical potential.
The P-P Coulomb potential, however, leads to a
cross section without Coulomb rainbow, very close to Rutherford's.
This absence of Coulomb rainbow
is due to the fact
that the P-P interaction does not account for the extension
of the colliding nuclei, and hence remains quite close to a
Coulomb potential between two pointlike particles.
On the contrary, the P-S calculation, taking account of the
finite charge distributions of the core and the target,
exhibits a stronger deviation from purely Rutherford scattering.

The inclusion of nuclear optical potentials
increases the deviation from the Rutherford cross section.
The rainbow pattern hence reflects mostly the nature of the
$P$-$T$ interactions and in particular its 
deviation from a pure Coulomb potential: the larger the deviation,
the smaller the angle at which the drop occurs.
Interestingly, it is not related to the loss of flux from
the elastic-scattering channel towards the breakup channel.
\tbl{t1} displays the total elastic-breakup cross sections $\sigma_{\rm bu}$
obtained in our calculations.
We observe that it actually decreases
when the deviation from the Rutherford cross section increases.
This counterintuitive result suggests that
loss from the elastic-scattering channel does not explain
the rainbow pattern.
A loss of flux is also possible towards
the absorption induced by the imaginary part of the optical
potentials, which simulates all inelastic processes but breakup.
The absorption cross section $\sigma_{\rm abs}$ is also given in \tbl{t1}.
Since both P-P and P-S potentials are real, they do not lead
to absorption.
Therefore, the transfer to this channel cannot explain the
rainbow pattern either,
confirming that it is mostly sensitive to the $P$-$T$ interactions.
Note that since elastic breakup and absorption include all the
inelastic processes simulated within the DEA,
the total reaction cross section reads
$\sigma_{\rm R}=\sigma_{\rm bu}+\sigma_{\rm abs}$.

\begin{table}
\center
\begin{tabular}{l|c|c|ccc}
Interaction         & P-P &  P-S & \multicolumn{3}{c}{Coulomb+nuclear}\\ \hline
$|E_0|$ (MeV)         & 0.5 & 0.5  & 0.5      & 0.05   & 5    \\ \hline
$\sigma_{\rm bu}$ (b)& 2.58& 2.10 & 1.70     & 23.57  & 0.07 \\
$\sigma_{\rm abs}$ (b)& 0  & 0    & 3.87     & 4.41   & 3.58  
\end{tabular}
\caption{Total breakup and absorption cross sections for three
$P$-$T$ interaction choices 
and various neutron separation energies.}\label{t1}
\end{table}

To evaluate the sensitivity of the elastic scattering
to the extension of the projectile wave function,
we perform calculations
using \ex{11}Be-like projectiles 
with more extended/compact hal\oe s.
This is achieved by varying the binding energy of the ground state
from $|E_0|=0.5$~MeV to 50~keV and 5~MeV.
The N sides of the corresponding elastic-scattering cross sections
are plotted in \fig{f2}.
We observe that the N side drops faster 
as the binding energy decreases, i.e. as the projectile gets more diffuse.
The frequency and amplitude of the oscillations
before the rainbow, however, remain constant.
The sensitivity of the slope of the drop to the projectile extension
remains limited, unlike the change in the breakup cross section
(see \tbl{t1}).
This reinforces our idea that breakup
does not directly affect elastic scattering.
The absorption, though, seems to evolve similarly
to the slope of the drop. However, since comparable results are
obtained with real nuclear $P$-$T$ potentials, we conclude that the
absorption and the slope of the drop have common roots,
i.e. the size of the projectile, but no causal relationship.
This study shows that the general
features of the elastic-scattering cross section
reflect mostly the nature of the $P$-$T$ interactions,
that they are slightly affected by the extension of the projectile,
but are not influenced by losses of flux towards other channels.

\section{Elastic breakup}
\begin{figure}
\center
\includegraphics[width=10cm]{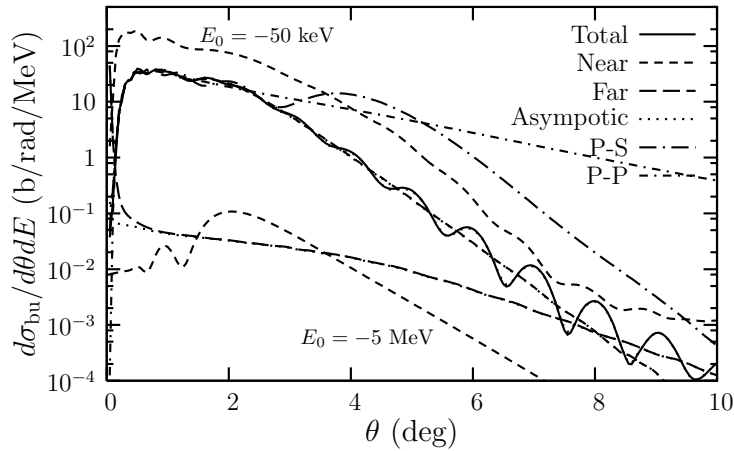}
\caption{N/F decomposition of the angular distribution for the elastic breakup
of \ex{11}Be on Pb at 69~MeV/nucleon for a \ex{10}Be-n relative
energy $E=0.5$~MeV.
The influence of the nature of the $P$-$T$ interactions 
and of the extension of the projectile wave function is also illustrated.
}\label{f3}
\end{figure}

We now study in a similar way the angular distribution
obtained for the elastic breakup of \ex{11}Be
on Pb at 69~MeV/nucleon.
In \fig{f3}, this cross section
is plotted for a relative energy $E=0.5$~MeV
between the \ex{10}Be core and the neutron after dissociation (full line).
Its behaviour is very similar to that of the elastic-scattering cross section:
It exhibits a first oscillatory pattern followed by a drop beyond $2^{\circ}$,
during which it presents a second set of oscillations.
This second set of oscillations is also explained as
interferences between the N and F sides.

The calculation of the N/F decomposition of the breakup
angular distribution requires a special treatment compared to the
elastic scattering. Indeed, both sides diverge at forward angles.
This unphysical result is due to the diverging behaviour
of the Hankel functions
$H_\mu$ for $\mu\ne0$ at small arguments \cite{AS70}
\beq
H_\mu^{\tiny \left(\!\!\!\!\begin{tabular}{c}
{1}\\
{2}
\end{tabular}
\!\!\!\!\right)}(z)\flim{z}{0}\mp\frac{i}{\pi}\Gamma(\mu)(z/2)^{-\mu}
\ \ \ \ \ \ \mbox{for } \mu>0.
\eeqn{e11}
This problem can be solved using the asymptotic expression
of the Hankel functions \eq{e7b} down to $0^{\circ}$ (dotted lines).
This seemingly rough approximation is validated by checking that
the angular distribution obtained with the asymptotic expression
of the Bessel function is identical to its exact expression.
This comparison also confirms the interpretation
of the N/F decomposition as positive and negative deflections
in elastic breakup.

We observe that the breakup process
is also dominated by the N side, and that the
oscillatory pattern observed at forward angles and the exponential drop 
of the cross section beyond $2^{\circ}$
are not due to N/F interferences.
This similarity with elastic scattering
suggests that the forward-angle features are due to
a process similar to a Coulomb rainbow.
To confirm this idea, we perform calculations with both
P-S and P-P Coulomb interactions
and display their N sides in \fig{f3}.
Again, the P-P calculation does not exhibit
any oscillation nor sudden drop, as already noted in the
DEA calculation of the Coulomb breakup of \ex{8}B (see Fig.~7 of \Ref{GCB07}).
The P-S calculation also exhibits a pattern
similar to the Coulomb rainbow observed in elastic scattering.
This illustrates that the $P$-$T$ interactions affect the angular 
distributions for both elastic scattering and breakup in a similar way,
suggesting that the breakup channel does not affect the
features of the elastic-scattering cross section.

To complete this study,
we repeat our calculations with extended/compacted
\ex{11}Be-like projectiles.
Their N sides, displayed in \fig{f3} (upper and lower dashed curves),
confirm the elastic-scattering results, that the slope of the
drop of the angular distribution is slightly
sensitive to the projectile extension.
Note that this conclusion holds although
the magnitude of the breakup cross section varies widely
with the binding of the projectile, as noted earlier.

\section{Conclusion}

In this work, we analyse the angular distribution for elastic
scattering and breakup using an extension of the N/F decomposition
\cite{CIH85,HM84} applied to the DEA that describes
coherently both processes \cite{BCG05,GBC06}.
As expected for Coulomb-dominated reactions, we show that
the collision of \ex{11}Be on Pb at 69~MeV/nucleon
is dominated by the N side, i.e. positive deflection,
and that N/F interferences are observed only at large scattering angles.
The Coulomb-rainbow features are shown to depend mostly on the nature of the $P$-$T$
interactions, and in particular their deviation from a purely Coulomb
potential. We also observe that the slope of the drop in the angular
distribution is sensitive to the extension of the projectile wave function.
These observations hold for both elastic scattering and breakup.

Examining both elastic scattering and breakup in parallel,
this work establishes that the loss of flux
towards inelastic channels, like breakup or absorption,
does not influence the general features mentioned above. 
In particular, the slope of the drop of the angular distribution
is related to the extension of the projectile, and not to
breakup or absorption.
Although this sensitivity may be too small to be used as a
probe of the halo structure, it can help understanding discrepancies
between theoretical predictions and experimental data.
In particular, it suggests
that the reduction of the cross section for elastic-scattering
of halo nuclei around the Coulomb barrier is not due to a loss of flux towards
the breakup channel, as proposed in Refs.~\cite{Dip04,Dip10}, but
is more a direct influence of the halo on the collision.
To precise this idea, we plan to apply the present methodology,
and in particular the parallel study of both elastic scattering and breakup,
to low-energy reactions within the
continuum discretised coupled channel (CDCC) model \cite{DBD10}.

\section*{Acknowledgments}
This text presents research results of the Belgian Research Initiative
on eXotic nuclei (BriX), programme P6/23 on interuniversity attraction
poles of the Belgian Federal Science Policy Office.
P.~C. acknowledges the support
of the NSF grant PHY-0800026.
He also thanks H.~Feldmeier for an interesting discussion on the subject.
M.~S.~H. is supported by the Brazilian agencies CNPq and FAPESP.

\end{document}